\documentstyle{resceu}
\journalid{98}{6}{1998}
\articleid{1}{13}

\def\beq{\begin{equation}}
\def\eeq{\end{equation}}
\def\beqar{\begin{eqnarray}}
\def\eeqar{\end{eqnarray}}

\def\fun#1#2{\lower3.6pt\vbox{\baselineskip0pt\lineskip.9pt
  \ialign{$\mathsurround=0pt#1\hfil##\hfil$\crcr#2\crcr\sim\crcr}}}

\def\b1#1{\hbox{${}^{1#1}$B}}
\def\c1#1{\hbox{${}^{1#1}$C}}
\def\o1#1{\hbox{${}^{1#1}$O}}
\def\ne2#1{\hbox{${}^{2#1}$Ne}}
\def\ti4#1{\hbox{${}^{4#1}$Ti}}

\title	{Axion Mass Limits from Cooling Neutron Stars}
 {Axion Mass Limits from Cooling Neutron Stars}

\author{Hideyuki UMEDA$^1$,  Naoki IWAMOTO$^2$ \\
 Sachiko TSURUTA$^3$, Letao QIN$^3$, \& Ken'ichi NOMOTO$^1$ \\
\\
{\it  $^1$Research Center for the Early Universe and Department of Astronomy,
}\\ {\it School of Science, University of Tokyo, Bunkyo-ku,
Tokyo 113-0033, Japan} \\ 
{\it  umeda@astron.s.u-tokyo.ac.jp}\\
{\it
$^2$ Department of Physics and Astronomy, University of Toledo, 
Toledo, Ohio 43606-3390, USA
}\\ 
{\it
$^3$ Department of Physics, Montana State University,
Bozeman, Montana 59717-0350, USA}\\ \\
{
Proceedings of the 
Symposium (17 - 20 November 1997, Tokyo)}\\ 
{``Neutron Stars and Pulsars'' }\\
{ eds. N. Shibazaki \etal
(World Scientific), p. 213, 1998
}\\ 
}	{Umeda, et al.}

\abstract{\noindent
The thermal evolution of a neutron star is studied by including the
energy loss due to axion emission. Two axion models and three types of
neutron-star matter equation of state are used with the effects of
nucleon superfluidity properly taken into account. In comparison with 
the observational data of PSR0656+14 from ROSAT, the upper limits on 
the axion mass are found to be $m_a < 0.06 - 0.3$ eV and 0.08$ -$ 0.8 eV 
for the KSVZ and DFSZ axion models, respectively, with the soft 
equation of state giving the most stringent limits.
}

\newcommand\etal{{\it et al. }}
\newcommand\re{\noindent}
\newcommand\half{\frac{1}{2}}
\input{psfig.sty}

\begin{document}

     The axion arises as a solution to the strong CP problem 
(Turner 1990, Raffelt 1990). While the standard axion model was 
excluded by experiments, the 
invisible axion model has survived 
mainly because the axion's coupling to matter is weak, which is an
unknown parameter in the theory. Over the years, various laboratory 
experiments as well as astrophysical arguments have been used to 
constrain its parameters. Since laboratory experiments can explore 
only a limited parameter regime, including those planned in the 
foreseeable future, astrophysical considerations have played an 
important role in placing the limits on the axion parameters.
Within these limits, the axion remains as one of the candidates for
dark matter. 

     There are two types of axion models---the 
KSVZ (hadronic) model (Kim 1979, Shifman \etal 1980)
and the 
DFSZ model (Dine \etal 1981, Zhitnitskii 1980). In the KSVZ model, the axion 
couples only to the photons and hadrons, while in the DFSZ model the 
axion couples to the charged leptons as well. The axion-fermion and 
axion-photon coupling constants as well as the axion mass are unknown 
parameters in these theories. 
Currently, cosmological arguments give $m_a > 10^{-5}$ eV (Abbott and
Sikivie 1983, Dine and Fischler 1983).
The limit from Supernova 1987A, which used to give $m_a < 10^{-3}$ eV,  
is now somewhat relaxed  $m_a < 0.01$ eV (Raffelt and Seckel 1991,
Janka \etal 1996). The red giant limit 
$m_a < 0.009/\cos^2\beta $ (Raffelt and Weiss 1995) applies 
only to the DFSZ model. The laboratory experiments give weaker limits.

     In the present paper we study how axion emission affects the 
thermal evolution of neutron stars. We use the neutron star 
evolutionary code with three types of equation of state to calculate 
the surface temperature of neutron stars. 
We compare theoretical cooling curves with observation and obtain
the upper limits on the axion mass, which are weaker than, but 
comparable with, the limit from SN 1987A.

\vspace{1pc}
{\bf Axion Emissivity:}
 In neutron stars, the dominant axion emission mechanisms are the following
bremsstrahlung processes in the stellar core: $n + n\rightarrow n+ n+a$, 
$p+p \rightarrow p+p+a$, and $n+p \rightarrow n+p+a$, where $n, p,$ and $a$
are the neutron, proton and axion. The energy loss rate of
each process, in the units  $\hbar = c=1$,
is given by (Iwamoto \etal 1998),
\begin{equation}
 \epsilon_{ann} = \frac{31 g_{ann}^2 }{3780\pi} {m_n^*}^2 p_F(n)
\left(\frac{f}{m_\pi}\right)^{4} F(x) (k_B T)^6,
\end{equation}
\begin{equation}
 \epsilon_{app} = \frac{31 g_{app}^2  }{3780\pi} {m_p^*}^2 p_F(p)
\left(\frac{f}{m_\pi}\right)^{4} F(y) (k_B T)^6,
\end{equation}
\begin{equation}
 \epsilon_{anp} \simeq \frac{31}{5670\pi} m_N^2 p_F(p)
\left(\frac{f}{m_\pi}\right)^{4}  G(x,y) (k_B T)^6,
\end{equation}
where 
\begin{equation}
F(z)\equiv 1-\frac{3}{2}z~\rm{arctan} \left(\frac{1}{z}\right) 
+\frac{z^2}{2(1+z^2)},
\end{equation}

\begin{eqnarray}
 G(x,y) &\equiv& \half (g^2+h^2)F(y)  \nonumber \\
& +& (g^2+ \half h^2) \Bigl[ F(\frac{2xy}{x+y}) 
+ F(\frac{2xy}{y-x}) \nonumber \\
&& +(\frac{y}{x})  \bigl\{ F(\frac{2xy}{x+y})-
F(\frac{2xy}{y-x})\bigr\} \Bigr] \nonumber  \\
  &+&(g^2+h^2)(1-y ~\rm{arctan}(1/y)), 
\end{eqnarray}
$g\equiv g_{app}+g_{ann}$, 
$h\equiv g_{app}-g_{ann}$; $x\equiv m_\pi /2p_F(n)$,
 $y\equiv m_\pi /2p_F(p)$; $f\simeq 1$ is the pion-nucleon 
coupling constant;
$p_F(n)\simeq 340(\rho/\rho_0)^{1/3}$
MeV/c, $p_F(p)\simeq 85(\rho/\rho_0)^{2/3}$ MeV/c are the nucleon
Fermi momenta; $m_p, m_n$ and $m_\pi$ are the proton and
neutron effective masses and pion mass, respectively. 
\begin{equation}
 g_{aii} \equiv \frac{c_i m_N}{(f_a/12)}
\end{equation}
is the axion-nucleon coupling constant, where $i=p$ (proton) or $n$ 
(neutron),  $m_N$ is the nucleon mass,
and $f_a$ is the axion decay constant. $c_i$ depends on the models: the 
DFSZ model gives
\begin{equation}
 c_p=-0.10-0.45 \cos^2\beta, ~c_n=-0.18+0.39 \cos^2\beta,
\end{equation}
and the KSVZ (hadronic) model gives 
\begin{equation}
c_p=-0.385, \quad c_n=-0.044.
\end{equation}
The axion mass is related to $f_a$ via 
\begin{equation}
m_a = \frac{0.0074}{f_a/(10^{10}{\rm GeV})} \rm{eV}.
\end{equation}
 We note that axion emission is suppressed 
if nucleons become superfluid, as in the case of neutrino emission
involving nucleons.


\begin{figure}[t]
\vskip 12.5cm
\hskip -.0cm
 \psfig{figure=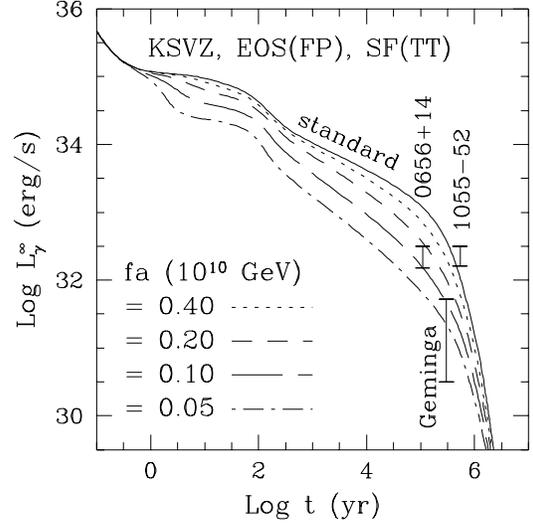,height=1.5in}
\vskip -9.cm
  \caption{Cooling curves with KSVZ axion emission } 
\end{figure}

\begin{figure}[h]
\vskip 12.5cm
\hskip -.0cm
 \psfig{figure=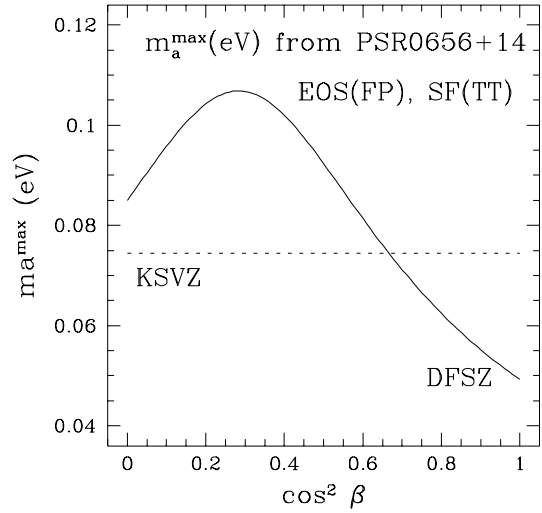,height=1.5in}
\vskip -9.cm
  \caption{Axion mass limits ($m_a^{max}$) 
in the FP model } 
\end{figure}

\vspace{1pc}
{\bf Results and Discussion:}
 We employ the numerical calculation code essentially the same as the
one described in Umeda \etal (1994), except for the inclusion of  the
energy loss due to axion emission. We neglect internal and other possible 
heating mechanisms as well as the existence of non-standard 
cooling mechanisms. The baryon mass of the neutron star
is set to 1.4 $M_\odot$.

 Theoretical cooling curves are compared with the
observational data for three pulsars: PSR 1055-52, Geminga and PSR 0656+14
(see Tsuruta 1998 and Becker 1994 for references).  
 The energy loss rate due to axion emission is proportional to the axion
mass squared, $m_a^2$; therefore, we can obtain the 
upper limit on the axion mass from the condition that the 
cooling curve does not pass below the
lower bounds on the observational points.

 In Figure 1, we show the {\it standard} cooling curve 
and those with the KSVZ axion model for four
different axion masses (or $f_a$).  The FP equation of state and the TT
neutron $^3P_2$ superfluid energy gap (Takatsuka and Tamagaki 1993)
are adopted. 
Since the data point for PSR 1055-52 is located above the standard
cooling curve, we do not use this data: this is likely to be due to
some other (unknown) effects.
Conservative limits can be obtained
by using the other two data. Figure 1 shows that the PSR 0656+14 
gives a more stringent limit than Geminga, and hence we obtain the axion 
mass limit from the lower bound on the PSR 0656+14 data.

 The results for both the KSVZ and DFSZ axion models with 
stiff (PS), medium (FP)  and
soft (BPS) equations of state are summarized in Figures 2-4. The BPS
model gives the more stringent limit. This is because the TT gap
vanishes in the high density region (i.e., inside the stellar core)
with this equation of
state; thus, axion emission is not suppressed. Extending superfluidity 
to higher density regions will have an effect similar to increasing the 
stiffness of the equation of state. For example, in the FP model, if the AO 
neutron $^3P_2$ gap (Amundsen and \O stgaard 1985) is adopted, 
$m_a^{\rm max}$ is 0.3 eV,
while if there is no neutron $^3P_2$ superfluid, $m_a^{\rm max}$ is 0.06 eV.
Note, however, that the AO model probably overestimates the energy
gap at high densities, 
because the density dependence of the neutron effective mass is neglected.
Future refinements of the observation will provide more stringent limits.

 This work is supported in part by the grant-in-Aid for Scientific
Research (05242102, 06233101, 6728) and COE research (07CE2002) of the
Ministry of Education, Science and Culture in Japan, and by the NASA
(NAGW-2208, NAG5-2557) and NSF (PHY-9722138).

\begin{figure}[t]
\vskip 12.5cm
\hskip -0cm
 \psfig{figure=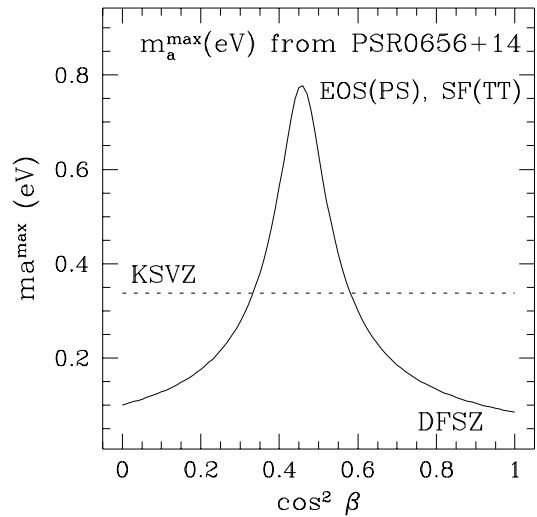,height=1.5in}
\vskip -9.cm
  \caption{Same as Fig. 2 in the PS model } 
\end{figure}

\begin{figure}[t]
\vskip 12.5cm
\hskip -0cm
 \psfig{figure=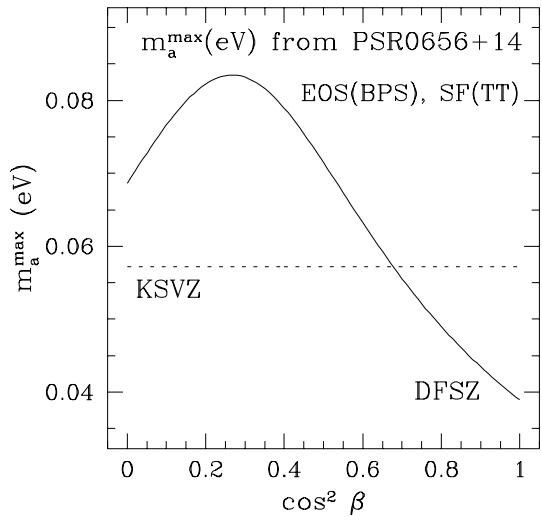,height=1.5in}
\vskip -9.cm
  \caption{Same as Fig. 2 in the BPS model } 
\end{figure}

\bigskip
{\bf References}
\vspace{1pc}

\re
Abbott, L. F., and Sikivie, P., 1983, Phys. Lett., B120, 133
\re
Amundsen, L., and \O stgaard, E., 1985, Nucl. Phys., A437, 487

\re
Becker, W., 1994, Ph.D. Thesis, M\"unchen University

\re
Dine, M., Fischler, W., and Srednicki, M., 1981, Phys. Lett., 104B 199

\re
Dine, M., and Fischler, W., 1983, Phys. Lett, B120, 137

\re
Iwamoto, N., Umeda, H., Tsuruta, S., Nomoto, K. and Qin, L., 
1998, in preparation

\re
Janka, H.-T., Keil, W., Raffelt, G., and Seckel, D., 1996,
Phys. Rev. Lett., 76, 2621 

\re
Kim, J. E., 1979, Phys. Rev. Lett., 43, 103

\re
Raffelt, G., 1990, Phys. Rep., 198, 1

\re
Raffelt, G., and Seckel, D., 1991, Phys. Rev. Lett., 67, 2605

\re
Raffelt, G., and Weiss, A,, 1995, Phys. Rev., D51, 1495

\re
Shifman, M., Vainshtein, A., and Zakharov, V., 1980, Nucl. Phys.,
B166, 493

\re
Takatsuka, T., and Tamagaki, R., 1993, Prog. Theor. Phys. Suppl.,
112, 27

\re
Tsuruta, S., 1998, Phys. Rep., 292, 1

\re
Turner, M. S., 1990, Phys. Rep., 197, 67

\re
Umeda, H., Tsuruta, S., and Nomoto, K., 1994, ApJ, 433, 256

\re
Zhitnitskii, A. P., 1980, Sov. J. Nucl. Phys., 31, 260

\end{document}